# An Axion Interpretation of the ANITA Events

A. Nicolaidis

Theoretical Physics Department
Aristotle University of Thessaloniki, Greece

**ABSTRACT**: We suggest that the unusual events observed by the ANITA experiment originate from axion particles traversing the Earth. Under the influence of the geomagnetic field, the axion may oscillate into a photon and vice-versa. To amplify the axion transition into photon, we consider that the phenomenon takes place at resonance, where the effective photon mass is equal to the axion mass. This requirement fixes the axion mass at 200 eV. An axion at this mass scale reproduces the cold dark matter scenario. If our interpretation prevails, with the help of axions we can establish an axion tomography of the Earth.

The Antarctic Impulsive Transient Antenna (ANITA) experiment has observed two air shower events with energy ~ 500 PeV emerging from the Earth with exit angles ~ 30° above the horizon [1,2]. The steep arrival angle implies that the candidate particle propagates a distance inside the Earth of the order of the Earth's radius $R_E$ (6371 km).

One might think the neutrino lies at the origin of the ANITA unusual events. The neutrinos interact with the nucleons through the weak charged current, resulting into absorption, and the weak neutral current, which implies a redistribution of the neutrino energy. For a detailed analysis see [3], where a Mellin transform of the neutrino transport equation provides the shadowing factor of ultrahigh high energy neutrinos. The neutrino-nucleon cross section rises with energy and at energies above a few TeV the Earth is becoming opaque to neutrinos. We are engaged then to search for a solution next or beyond the Standard Model. In the present work we examine the prospect that the axion particle might serve this purpose.

Let us recall that the raison d'être of the axion particle is the strong CP problem. The QCD Lagrangian respects all symmetries (P, C, CP…). At low energies, the non-linear nature of the theory introduces a non-trivial vacuum which violates the CP symmetry. The CP-violating term is parameterized by $\theta$ and experimental bounds indicate that $\theta < 10^{-9}$. This is a very small number, and the smallness of this parameter creates what is known as the strong CP problem. An elegant solution has been offered by Peccei-Quinn [4]. A global $U(1)_{PQ}$ symmetry is introduced, the spontaneous breaking of which provides the cancellation of the $\theta$-term. As a byproduct, we obtain the axion field, the Nambu-Goldstone boson of the broken $U(1)_{PQ}$ symmetry. There are extensive reviews covering the theoretical aspects and the experimental searches for the axion [5-7].

A general feature of the axion is its two-photon coupling

$$L_{a_{\gamma\gamma}} = -\frac{1}{4} g a F_{\mu\nu} \tilde{F}^{\mu\nu} = g a \vec{E} \cdot \vec{B} \qquad (1)$$

where $a$ is the axion field, $F_{\mu\nu}(\tilde{F}^{\mu\nu})$ the (dual) electromagnetic field strength tensor and $g$ the photon-axion coupling constant. Accordingly, in the presence of a magnetic field B, a photon may oscillate into an axion and vice-versa. A prototype experiment in the search for solar axions is the CAST experiment, which set the limit $g < 10^{-10}$ GeV$^{-1}$ [8, 9]. The CAST experiment involves a magnetic field B = 9 T and a magnetized region $L$ = 9.3 m. Therefore, the relevant scale $(BL)^2$ is $(BL)^2 \approx 7000$ T$^2$m$^2$. Our proposal involves Earth's magnetic field, a magnetic dipole with a mean value $B_0 \approx 3 \times 10^{-5}$ T on the Earth's surface. The weakness of the geomagnetic field B is compensated by the larger $L$ value, of the order of Earth's radius $R_E$. Therefore, in our case the scale is $(BL)^2 \approx 36100$ T$^2$m$^2$. This increased value allows a higher accuracy and the exploration of a new range of $g$ and $m_a$ (coupling constant and axion mass respectively) [10, 11].

Consider a travelling photon of energy $E$ and let us define as z-axis the direction of photon's propagation. The polarization of the photon $\vec{A}$ lies then at the x–y plane. The photon is moving in the presence of the geomagnetic field $\vec{B}$. The component of $\vec{B}$ parallel to the direction of motion does not induce photon-axion mixing. Following eq. (1), the transverse magnetic field $\vec{B}_T$ couples to $A_{\parallel}$, the photon polarization parallel to $\vec{B}_T$ and decouples from $A_{\perp}$, the photon polarization orthogonal to $\vec{B}_T$. The photon-axion mixing is governed by the following equation:

$$\left(E - i\partial_z + \mathbf{M}\right)\begin{pmatrix} A_{\parallel} \\ a \end{pmatrix} = 0 \qquad (2)$$

The 2-dimensional matrix **M** is

$$\mathbf{M} = \begin{pmatrix} -\dfrac{m_\gamma^2}{2E} & \dfrac{g B_T}{2} \\ \dfrac{g B_T}{2} & -\dfrac{m_a^2}{2E} \end{pmatrix} \qquad (3)$$

For a photon, moving in a medium with number density of electrons $N_e$, the effective photon mass $m_\gamma$ is given by

$$m_\gamma^2 = \frac{4\pi\alpha N_e}{m_e} \tag{4}$$

Assuming that Earth's material contains an equal number of protons and neutrons, we obtain the estimate [12, 13]

$$N_e \sim \rho/(2m_N) \tag{5}$$

Therefore

$$m_\gamma^2 = \frac{2\pi\alpha}{m_e m_N}\rho \tag{6}$$

The density of the Earth as a function of the distance is rather well known and very close to the two-density model description, in which the core and the mantle each have a separate and constant energy [14]. We take the core of the Earth to be a sphere whose radius is $R_2 = 3490$ km and whose constant density is 11.0 g/cm$^3$. The mantle, a spherically symmetric shell of constant density 4.4 g/cm$^3$, surrounds the core and extends out to $R_E = 6371$ km. For a nadir angle $\sim 60°$, our particle crosses the Earth at a distance 5517 km far from the center, traversing a distance of 6371 km within the Earth. Moving entirely within the mantle, $\rho = 4.4$ g/cm$^3$ and the effective photon mass is $m_\gamma \sim 200$ eV.

Matrix **M** is diagonalized through the angle $\Theta$ with

$$\tan 2\Theta = \frac{2gB_T E}{m_a^2 - m_\gamma^2} \tag{8}$$

Defining

$$D = \frac{1}{2E}\left[(m_a^2 - m_\gamma^2)^2 + 4g^2 B_T^2 E^2\right]^{1/2} \tag{9}$$

$$\sin 2\Theta = \frac{gB_T}{D} \tag{10}$$

we obtain for the probability that an axion converts into a photon after travelling a distance $s$

$$P(a \to \gamma) = \sin^2 2\Theta \, \sin^2 \frac{Ds}{2} \qquad (11)$$

A resonance phenomenon occurs, offering the maximum probability, when

$$m_a = m_\gamma \qquad (12)$$

We gather that the most favorable value for the axion mass is $m_a \sim 200$ eV. Proceeding along these lines we obtain that at resonance and for values of $Ds \ll 1$

$$P(a \to \gamma) = \frac{1}{4} g^2 B_T^2 s^2 \qquad (13)$$

Putting the appropriate numbers ($g = 10^{-10}$ GeV$^{-1}$, B=3 × 10$^{-5}$ T, $s = R_E$) we obtain

$$P(\alpha \to \gamma) = 10^{-16} \qquad (14)$$

Let us summarize our findings. We suggest that highly energetic axions traverse the Earth and they are becoming photons under the influence of the geomagnetic field. These photons create the showers observed by ANITA. The proposed mechanism suggests a mass scale for the axion at 200 eV. The physical properties of the QCD axion are to large extent determined by the scale $f_a$ of the PQ symmetry breaking, similar to how the low energy pion interactions are fixed by the pion decay constant $f_\pi$. Next to QCD interactions, we should include the electroweak interactions and also the gravitational interactions [15]. Thus, the obtained mass scale of 200 eV is not unnatural. A proposed experiment [16] is dedicated to explore axions in a mass range around several eV. What is most interesting is that our axion can solve also the dark matter issue. Cosmological N-body simulations with dark matter indicate that an axion with a mass above a hundred eV will provide power spectra almost indistinguishable from ΛCDM [17, 18]. Thus two problems disappear with a single suggestion.

One might wonder what the origin of these energetic axions is. We can imagine that the inverse phenomenon takes place at gigantic extragalactic scale. VHE photons in the presence of magnetic fields at their source suffer conversion into axions, thus avoiding absorption by γγ collisions on the extragalactic background light. Through this mechanism we obtain a spectrum of "hard" photons and axions [19]. These axions may reach our planet. If our model prevails, then these axions crossing the Earth may be useful in order to establish an axion tomography of the Earth. Notice that a neutrino tomography of the Earth has been already achieved [12, 13, 20]. On the other hand, a fraction of these axions may be converted into photons in the Milky Way. These VHE photons should be of prime interest to the CTA experiment [21]. A multimessenger exploration of space and particle physics is opened.

There are other proposals to address the unusual ANITA events. It has been suggested that an axion pulse is transformed into an electromagnetic pulse in Earth's ionosphere [22]. Subsequently the down-going radio wave is reflected in the Antarctic ice, giving rise to the peculiar events. In another direction a supersymmetric interpretation has been advanced to explain the ANITA events [23]. Clearly we need more data to unravel the underlying mechanism.

**Acknowledgement** Dimitris Evangelinos of the Physics Department helped enormously in the typesetting of the text.